\def\e{{\rm e}}
\def\del{\partial}
\def\half{{1\over2}}

\def\abs#1{{\left|{#1}\right|}}
\def\vev#1{\langle #1 \rangle}

\def\del{\partial}
\def\dslash{\del\kern-0.55em\raise 0.14ex\hbox{/}}
\def\Lag{{\cal L}}

\newcommand{\PRD}[3]{Phys. Rev. {\bf D{#1}}, {#3} (19{#2})}

\newcommand{\NPB}[3]{Nucl. Phys. {\bf B{#1}}, {#3} (19{#2})} 
\newcommand{\PLB}[3]{Phys. Lett. {\bf B{#1}}, {#3} (19{#2})}

\newcommand{\PTP}[3]{Prog. Theor. Phys. {\bf {#1}}, {#3} (19{#2})} 
 
\newcommand{\ANN}[3]{Ann. Phys. {\bf {#1}}, {#3} (19{#2})}

\newcommand{\MPL}[3]{Mod. Phys. Lett. {\bf A{#1}}, {#3} (19{#2})}

\newcommand{\hmu}{\hat\mu}
\newcommand{\hnu}{\hat\nu}

\documentclass[12pt]{article}
\usepackage{epsf}
\begin{document}
\title{Dynamics of Nonintegrable Phases \\in \\
Softly Broken Supersymmetric Gauge Theory \\with\\ Massless Adjoint Matter}
\author{Kazunori Takenaga \vspace{1cm}$^{}$\thanks {Email address: 
takenaga@synge.stp.dias.ie}\\ 
\it {School of Theoretical Physics,}\\ 
{\it Dublin Institute for Advanced Studies, 10 Burlington Road, Dublin 4,}\\
{\it Ireland}}
\date{} 
\maketitle
\baselineskip=18pt
\vskip 2cm
\begin{abstract}
We study $SU(N)$ supersymmetric Yang-Mills theory with massless
adjoint matter defined on $M^3\otimes S^1$. 
The $SU(N)$ gauge symmetry is broken maximally 
to $U(1)^{N-1}$, independent of the number of flavor and
the boundary conditions of the fields associated with the 
Scherk-Schwarz mechanism of supersymmetry breaking.
The mass of the Higgs scalar is generated through
quantum corrections in the extra dimensions.
The quantum correction can become manifest by a finite 
Higgs boson mass at low energies even in the limit of 
small extra dimensions thanks to the supersymmetry breaking 
parameter of the Scherk-Schwarz mechanism.
\end{abstract}
\vskip 2cm
\begin{flushleft}
DIAS-STP-01-05\\
May 2001\\
\end{flushleft}
\addtolength{\parindent}{2pt}
\newpage
\section{Introduction}
The dynamics of nonintegrable phases is one of the most important phenomena
when one studies (supersymmetric) gauge 
theory in a space-time where
one of the space coordinates is compactified on a 
topological manifold \cite{hosotania}. 
The dynamics are caused essentially by quantum effects in extra 
dimensions, reflecting the topology of the extra dimensions. 
\par
Component gauge fields for compactified directions
can develop vacuum expectation values. 
The vacuum expectation values, which correspond to the constant 
background fields, are
also related with the eigenvalues 
of the Wilson line integrals for the compactified direction. 
Therefore, they are dynamical variables 
and cannot be gauged away. By studying the effective potential for 
the phases, in perturbation theory, for example, one can investigate how gauge 
symmetry is broken dynamically.
This shows that the quantum effects in the extra dimensions are remarkable
and should be taken into account when we study the theory. 
\par
The dynamics of the nonintegrable phases have been studied 
extensively\cite{davies, mac, higuchi} since a pioneering 
work \cite{hosotania}. 
In nonsupersymmetric gauge theories, one
can, in principle, determine how the gauge symmetry is broken dynamically.
It has been known that the gauge symmetry breaking patterns depend on
the number, the boundary conditions of the fields and the representation 
under the gauge group of matter \cite{hohosotani, ho, hatanaka}.
On the other hand, in supersymmetric
gauge theories, the effective potential for the phases vanishes 
as long as supersymmetry is not broken. 
This is because contributions coming from bosons and fermions
in a supermultiplet to the constant 
background field
cancel each other. One natural way to break supersymmetry
is to resort to the Scherk-Schwarz mechanism \cite{ss, fi}.
\par
Supersymmetry is broken by the boundary conditions of the 
fields for compactified
directions in the Scherk-Schwarz mechanism. By using symmetry degrees of 
freedom of the theory, one can twist the boundary conditions 
of the field in such a way that they are different between bosons and 
fermions in a supermultiplet. The boundary condition associated 
with the $U(1)_R$ symmetry is a candidate and breaks supersymmetry 
softly \cite{ss, fi, takenaga}. And it has been 
also pointed out that the supersymmetry breaking terms
resulting from the mechanism have attractive features such as 
flavor blindness and only two 
parameters \cite{takenaga, takenagab}. Once 
the supersymmetry is broken, we obtain nonvanishing 
effective potential for the nonintegrable phases in perturbation theory
and can discuss how the gauge symmetry is broken dynamically.
\par
In a previous paper \cite{takenagab}, the 
author studied the dynamics of the nonintegrable
phases in the softly broken supersymmetric gauge theories with matter 
defined on $M^3\otimes S^1$, where $M^3, S^1$ are three-dimensional 
Minkowski space-time and a circle, respectively. 
In that paper the gauge group was assumed to be $SU(2)$. 
This paper is a generalization of the previous work. We 
shall consider the $SU(N)$ gauge group and  
study the dynamics of the nonintegrable phases in a model of
$SU(N)$ supersymmetric Yang-Mills theory with  
$N_F$ numbers of massless adjoint matter. We resort to the 
Scherk-Schwarz mechanism to break the supersymmetry
in the model.
\par
In nonsupersymmetric gauge theory with massless adjoint matter, the 
gauge symmetry breaking 
patterns are rich, depending on the values of the boundary 
conditions of the matter field \cite{hohosotani}. 
On the contrary, the $SU(N)$ gauge 
symmetry is broken dynamically to its maximal 
commutative subgroup, {\it i.e.}, $U(1)^{N-1}$
in our model. We have an unique gauge symmetry breaking pattern.
It is remarkable that this does not 
depend on the boundary 
conditions of the fields associated with the Scherk-Schwarz mechanism 
of supersymmetry breaking.  
\par
The component gauge field for the compactified direction acquires mass
through the quantum correction in the extra dimensions and becomes
a Higgs scalar in the adjoint representation under the $SU(N)$
gauge group at low energies \cite{hosotania}.
We obtain the mass of the Higgs
scalar in our model. The mass
explicitly depends on the gauge coupling, supersymmetry breaking 
parameter, and the number of flavor and suffers from a correction of
the compactification scale. 
\par
The mass spectrum of the model at 
low energies is also obtained, and we discuss low-energy
effective theory.
If we take the limit of small extra dimensions, all the effects of the
extra dimensions whose mass scales are given by the compactification
scale are decoupled
from low-energy physics. It implies that the effect of the extra
dimensions never appears at low energies. The nontrivial limit of 
small extra dimensions, however, is possible thanks to
the supersymmetry breaking parameter of the Scherk-Schwarz mechanism.
The effect of the extra dimensions can become manifest by 
the Higgs scalar as having finite mass at low energies.  
\par
In the next section we shall start with reviewing
briefly the effective potential for the
nonintegrable phases. In order to make discussions clearer and carry out 
analytic calculations, we take our space-time to be $M^3\otimes S^1$ as 
in the previous work.
In section $3$ we shall 
discuss the gauge symmetry breaking through the dynamics of the
nonintegrable phases. We obtain the mass spectrum
in three dimensions and discuss the low-energy
effective theory in the limit of small extra dimensions in 
section $4$. Conclusions are given in section $5$.
\section{Effective Potential for Nonintegrable Phases}
Let us first consider, $SU(N)$ supersymmetric Yang-Mills theory
defined on $M^3\otimes S^1$, where $M^3$
is three-dimensional Minkowski space-time and $S^1$ is a circle. 
The coordinates of $M^3$ and
$S^1$ are denoted by $x^{\mu}$ and $y$, respectively. $x^{\hmu}$ stands for 
$x^{\hmu}=(x^{\mu}, y)$ and $L$ is the length of the 
circumference of $S^1$. 
\par
The on-shell Lagrangian is given by
\begin{equation}
\Lag =\mbox{tr}\Bigl[-\half F_{\hmu\hnu}F^{\hmu\hnu}
-i\lambda\sigma^{\hmu}D_{\hmu}{\bar\lambda}+
iD_{\hmu}\lambda\sigma^{\hmu}{\bar\lambda}\Bigr].
\end{equation}
Here $\lambda$ is 
the gaugino, the superpartner of the gauge boson $A_{\hmu}$.
The covariant derivative and field strength are defined by
\begin{equation}
F_{\hmu\hnu}=\del_{\hmu}A_{\hnu}-\del_{\hnu}A_{\hmu}-ig
[A_{\hmu}~~A_{\hnu}],\qquad 
D_{\hmu}\lambda=\del_{\hmu}\lambda -ig[A_{\hmu}~~\lambda],
\end{equation} 
respectively.
Here $g$ is the 
gauge coupling constant, and we normalize generators of $SU(N)$ as
$\mbox{tr}(T^aT^b)=\delta_{ab}/2$, where $a, b=1, \cdots, N^2-1$.
\par
It has been pointed out that the
component gauge field for the $S^1$ direction, denoted 
by $A_3\equiv \Phi$, can 
develop vacuum expectation 
values, reflecting the topology of $S^1$\cite{hosotania}. 
One can parametrize the vacuum expectation value as
\begin{equation}
\vev{\Phi}={1\over gL}\mbox{diag}(\theta_1, \theta_2, \cdots, \theta_N)
\quad \mbox{with}\quad \sum_{i=1}^{N}\theta_i=0.
\label{background}
\end{equation}
The constant background (\ref{background}) is also 
equivalent to introducing the nontrivial Wilson line integral 
for the $S^1$ direction 
\begin{equation}
W_c\equiv {\cal P}\e^{-ig\oint_{S^1}dy~\vev{A_3}}
=\left(\begin{array}{cccc}
\e^{-i\theta_1}& & & \\
&\e^{-i\theta_2} & &\\
& & \ddots & \\
& & &\e^{-i\theta_N}\end{array}\right),\quad (\theta_i~~(\mbox{mod}~~2\pi)).
\end{equation}
The gauge symmetry is broken in the Cartan subgroup
of $SU(N)$ by nontrivial 
values of $\theta_i$. The residual gauge symmetry is generated by 
generators commuting with $W_c$, {\it i.e.} $[W_c,~T^a]=0$. 
The phase $\theta_i$ is
called the nonintegrable phase \cite{hosotania}.
\par
If we expand the fields around 
the vacuum expectation value and integrate out 
the fluctuating fields up to the quadratic terms, we obtain 
the effective potential for the
nonintegrable phases in a one-loop approximation.
As we have noted 
in the introduction, we need to break supersymmetry to obtain 
nonvanishing effective potential at least in perturbation theory.
We shall resort 
to the Scherk-Schwarz mechanism \cite{ss, fi}.
According to the mechanism, supersymmetry 
is broken by the nontrivial boundary conditions of the
gaugino field $\lambda$ for the $S^1$ direction.
By using the $U(1)_R$-symmetry degrees of freedom in the theory, one 
can impose the boundary conditions on the field, which are 
defined by \cite{ss, fi, takenaga}
\begin{equation}
\lambda(x^{\mu}, y+L)=\e^{i\beta}\lambda(x^{\mu}, y).
\label{bcgaugino}
\end{equation}
The gauge field $A_{\hmu}$ satisfies the periodic boundary 
conditions. Thus, $A_{\hmu}$ and $\lambda$ obey the different boundary
conditions, so that supersymmetry is broken by the mechanism. 
\par
The effective potential for the nonintegrable 
phases is calculated, in the Feynman gauge, as \cite{takenagab}
\begin{equation}
V_{SYM}(\theta)={-2\over{\pi^2 L^4}}\sum_{n=1}^{\infty}\sum_{i, j=1}^{N}
{1\over n^4}\biggl(\cos[n(\theta_i-\theta_j)]-
\cos[n(\theta_i-\theta_j-\beta)]\biggr).
\label{sympot}
\end{equation}
The first and second terms in the potential come from the gauge boson
and gaugino, respectively. 
As we can see, if we take $\beta=0$, the supersymmetry is restored
to yield vanishing effective potential.
We can also recast the potential as   
\begin{equation}
V_{SYM}(\theta)={-2\over{\pi^2 L^4}}\sum_{n=1}^{\infty}
{1\over n^4}(1-\cos(n\beta))
\biggl(N+2\sum_{1\leq i<j\leq N}\cos[n(\theta_i-\theta_j)]\biggr).
\label{sympotanother}
\end{equation}
The leading term $N$ comes from the diagonal part with 
respect to $i$ and $j$.
It is important to note that the nontrivial 
phase $\beta$, which has a role in breaking supersymmetry, does 
not affect the location of absolute minima of the potential 
though the potential energy at the minimum depends on the phase. 
\par
Let us introduce $N_F$ numbers of massless 
adjoint matter into the theory. The chiral superfield
for the matter belongs to the adjoint representation 
under $SU(N)$, and the on-shell degrees of freedom 
in the chiral superfield are 
a complex scalar (squark) $\phi$ and a two-component Weyl
spinor (quark) $q$. We ignore the flavor index. 
We impose the boundary conditions associated with the $U(1)_R$
symmetry on the squark field\footnote{Strictly speaking, one has to
consider massive adjoint matter in order to have  
the boundary conditions associated with the $U(1)_R$ symmetry. 
The discussion here corresponds to the massless limit.}, 
\begin{equation}
\phi(x^{\mu}, y+L)=\e^{i\beta}\phi(x^{\mu}, y).
\label{bcmatter}
\end{equation}
The phase $\beta$ is common to all flavors as noted 
in Refs.\cite{takenaga} and \cite{takenagab}.
\par
By expanding the fields around the background (\ref{background}), we obtain
the effective potential for the nonintegrable phases coming from
the massless adjoint matter
\begin{equation}
V_{ADJ}(\theta)={{2N_F}\over{\pi^2 L^4}}\sum_{n=1}^{\infty}\sum_{i, j=1}^{N}
{1\over n^4}\biggl(\cos[n(\theta_i-\theta_j)]-
\cos[n(\theta_i-\theta_j-\beta)]\biggr).
\label{adjpot}
\end{equation}
Here $2N_F$ accounts for the on-shell degrees of freedom for the
massless adjoint matter. The first and second terms come from 
the quark and squark, respectively. 
Hence, the effective potential for the supersymmetric 
Yang-Mills theory with $N_F$ numbers of massless adjoint matter is given by
\begin{eqnarray}
V(\theta)&\equiv& V_{SYM}(\theta)+V_{ADJ}(\theta)\nonumber\\
&=&{{(2N_F-2)}\over{\pi^2 L^4}}\sum_{n=1}^{\infty}\sum_{i, j=1}^{N}
{1\over n^4}\biggl(\cos[n(\theta_i-\theta_j)]-
\cos[n(\theta_i-\theta_j-\beta)]\biggr)\nonumber\\
&=&{{(2N_F-2)}\over{\pi^2 L^4}}\sum_{n=1}^{\infty}
{1\over n^4}(1-\cos(n\beta))\biggl(N+
2\sum_{1\leq i<j\leq N}\cos[n(\theta_i-\theta_j)]\biggr)
\label{totpot}
\end{eqnarray}
Let us note, again, that the nontrivial phase $\beta$
does not affect the location of the absolute
minima of the potential.
\par
We immediately see that if $N_F=1$, the potential vanishes for any values 
of $\beta$. This is because the supersymmetric Yang-Mills 
theory with one massless adjoint
matter is nothing but ${\cal N}=2$ supersymmetric gauge theory
in four dimensions, and 
${\cal N}=1$ supersymmetry still remains even after imposing the
boundary conditions (\ref{bcgaugino}) and (\ref{bcmatter}) 
associated with the $U(1)_R$ symmetry. 
This, however, implies partial 
supersymmetry breaking through the boundary conditions associated with 
the $U(1)_R$ symmetry. In order to have 
nonvanishing effective potential for this case, one needs 
to impose the boundary conditions
associated with the $SU(2)_R$ [or $U(1)_J$ in ${\cal N}=1$ language] symmetry 
in addition to the $U(1)_R$. 
\section{Gauge Symmetry Breaking}
Let us discuss the gauge symmetry breaking through
the dynamics of the nonintegrable phases. We first study the effective 
potential (\ref{sympotanother}) for the case of the supersymmetric 
Yang-Mills theory. The potential is minimized when  
\begin{equation}
\theta_i-\theta_j=0.
\label{ymsol}
\end{equation}
Since $\sum_{i=1}^{N}\theta_i=0$, we have 
\begin{equation}
\theta_i={{2\pi m}\over {N}} ~~(m=0, 1, \cdots, N-1).
\end{equation}
It gives $\e^{i\theta_i}=\e^{2\pi i m/N}$, so that $W_c$ is an element of
the center of $SU(N)$ and commutes with all the generators 
of $SU(N)$. Therefore, the gauge symmetry is not broken in this theory.
This is the same result as in the case for the Yang-Mills
theory \cite{hosotania}. The potential energy 
at the minimum (\ref{ymsol}) is calculated as
\begin{equation}
V_{SYM}={{-N(N+1)}\over{\pi^2L^4}}\times
{{\beta^2(\beta-2\pi)^2}\over 48},
\end{equation} 
where we have used the formula
\begin{equation} 
\sum_{n=1}^{\infty}{1\over n^4}\cos(nt)
=-{1\over {48}}t^2(t-2\pi)^2 + {{\pi^4}\over{90}}\qquad 
(0\leq t\leq 2\pi).
\label{formula}
\end{equation}
The potential energy at the minimum is 
negative\footnote{The contribution from matter to the potential energy 
can make the potential energy positive as we will see later.
There may be a possibility to have zero
energy by adding an appropriate amount of matter.}.
This reflects the fact that 
the supersymmetry breaking of the Scherk-Schwarz mechanism is 
not a spontaneous breaking of supersymmetry, but an explicit breaking 
in our model. 
\par
Let us next study the gauge symmetry breaking in $SU(N)$ 
supersymmetric Yang-Mills theory with
$N_F$ numbers of massless adjoint matter. The effective potential is given 
by Eq. (\ref{totpot}). It may be convenient to recast it as
\begin{eqnarray}
V({\theta})&=&{{(2N_F-2)}\over {\pi^2L^4}}
\sum_{n=1}^{\infty}{1\over n^4}
\Biggl(N\Bigl(1-\cos(n\beta)\Bigr)\nonumber\\
&+&\sum_{1\leq i<j\leq N-1}
2\cos[n(\theta_i-\theta_j)]
-\cos[n(\theta_i-\theta_j-\beta)]
-\cos[n(\theta_i-\theta_j+\beta)]\nonumber\\
&+&2\cos[n(2\theta_1+\theta_2+\cdots+\theta_{N-1})]\nonumber\\
&-&\cos[n(2\theta_1+\theta_2+\cdots+\theta_{N-1}-\beta)]
-\cos[n(2\theta_1+\theta_2+\cdots+\theta_{N-1}+\beta)]\nonumber\\
&+&\cdots\nonumber\\
&+&2\cos[n(\theta_1+\theta_2+\cdots+\cdots+2\theta_{N-1})]\nonumber\\
&-&\cos[n(\theta_1+\theta_2+\cdots+2\theta_{N-1}-\beta)]
-\cos[n(\theta_1+\theta_2+\cdots+2\theta_{N-1}+\beta)]\Biggr),
\label{potanother}
\end{eqnarray}
where we have used $\theta_N=-\sum_{i=1}^{N-1}\theta_i$.
It is important to note that the potential is 
invariant\footnote{The effective potential is also invariant 
under $\beta\rightarrow\beta+2\pi im, m\in {\bf Z}$. This is traced back to
$\lambda\rightarrow \e^{2\pi m i}\lambda$. Note that physical 
region of $\beta$ is given by $-\pi\leq \beta \leq \pi$, except $\beta=0$.}
under
\begin{equation}
\theta_i \leftrightarrow \theta_j\quad (i\neq j),\quad
\beta    \leftrightarrow  -\abs{\beta}.
\label{symmetry}
\end{equation}
Classifications depending on the sign of $\theta_i-\theta_j-\beta$ are
necessary when we apply the formula (\ref{formula}) to the 
effective potential.
Thanks to the symmetries (\ref{symmetry}), however, the region 
given by $\theta_i-\theta_j\leq \beta$
is equivalent to that given by $\theta_i-\theta_j\geq \abs{\beta}$. 
It is enough for us take only the region of $\theta_i-\theta_j\geq \beta$  
into account. It also 
follows that the region $0<\beta\leq \pi$ 
is enough to study the potential.
\par
After straightforward calculations, we arrive at the
expression given by
\begin{eqnarray}
V(\theta)&=&{{(2N_F-2)}\over{\pi^2L^4}}~\beta^2
\biggl[{N\over {48}}(\beta-2\pi)^2+{{N(N-1)}\over{48}}(\beta^2+4\pi^2)
\nonumber\\
&+&{{2N}\over 4}
\biggl(\sum_{i=1}^{N-1}\theta_i^2 
+\sum_{1\leq i < j\leq N-1}\theta_i\theta_j\biggr)
-{{2\pi}\over 2}\sum_{i=1}^{N-1}(N-i)\theta_i
\biggr].
\label{potformula}
\end{eqnarray}
The absolute minima of the potential is given by solving 
$\del V(\theta)/\del\theta_k =0 (k=1, 2, \cdots, N-1)$, which
is read as
\begin{equation}
{N\over 2}
\biggl(\sum_{i=1}^{N-1}2\theta_i\delta_{ik}
+\sum_{1\leq i<j \leq N-1}(\theta_j\delta_{ik}+\theta_i\delta_{jk})\biggr)
=\pi\sum_{i=1}^{N-1}(N-i)\delta_{ik}, \quad k=1, 2, \cdots, N-1.
\label{extrem}
\end{equation}
This can also be written in the form 
\begin{equation}
{N\over 2} \left(\begin{array}{ccccc}
2&1&\cdots&\cdots& 1\\
1&2& & &\vdots\\
\vdots& &\ddots & &\vdots\\
\vdots& & & \ddots&\vdots\\
1&\cdots&\cdots&\cdots&2\\
\end{array}\right)\left(\begin{array}{c}
\theta_1\\
\theta_2\\
\theta_3\\
\vdots\\
\theta_{N-2}\\
\theta_{N-1}
\end{array}\right)=\pi
\left(\begin{array}{c}
N-1\\
N-2\\
N-3\\
\vdots\\
2\\
1\end{array}\right).
\label{stationary}
\end{equation}
All the (off-)diagonal elements of the matrix in Eq. (\ref{stationary}) are
$2(1)$.
The inverse of the matrix is given by
\begin{equation}
{1\over N}\left(\begin{array}{ccccc}
N-1&-1&\cdots&\cdots& -1\\
-1&N-1& & &\vdots\\
\vdots& &\ddots & &\vdots\\
\vdots& & & \ddots&\vdots\\
-1&\cdots&\cdots&\cdots&N-1\\
\end{array}\right),
\end{equation}
where all the (off-)diagonal elements of the matrix are $N-1(-1)$.
Therefore, the solution to Eq. (\ref{extrem}) is 
\begin{equation}
\left(\begin{array}{c}
\theta_1\\
\theta_2\\
\theta_3\\
\vdots\\
\theta_{N-2}\\
\theta_{N-1}
\end{array}\right)={{\pi}\over N}
\left(\begin{array}{c}
N-1\\
N-3\\
N-5\\
\vdots\\
-(N-5)\\
-(N-3)
\end{array}\right)\quad \mbox{or}\quad \theta_i={\pi\over N}
\Bigl(N-(2i-1)\Bigr)
\quad (i=1,\cdots, N-1).
\label{sol}
\end{equation}
Let us note that $\theta_N(=-\sum_{i=1}^{N-1}\theta_i)=-\pi(N-1)/N$.
We have found that the absolute minimum of the potential is 
located at
\begin{equation}
\vev{\Phi}={\pi\over{gL}}\mbox{diag}\biggl(
{{N-1}\over N}, {{N-3}\over N}, \cdots, 0,
{-{(N-3)}\over N}, {-{(N-1)}\over N}\biggr)\quad (\mbox{mod}~~2\pi),
\label{config}
\end{equation}
where $\theta_{i=(N+1)/2}=0$. 
\par
As an example, let us present results for the cases 
of $SU(2)\cite{takenagab}, SU(3)$, and $SU(5)$:
\begin{eqnarray}
\vev{\Phi} 
&=&{\pi\over{gL}}\mbox{diag}\biggl(\half, -\half\biggr) + \mbox{permutations}
\quad\mbox{for}\quad SU(2),
\nonumber\\
\vev{\Phi} &=&{\pi\over{gL}}\mbox{diag}\biggl({2\over 3}, 0, -{2\over 3}\biggr)
+\mbox{permutations}\quad\mbox{for}
\quad SU(3),\nonumber\\
\vev{\Phi} &=&{\pi\over{gL}}\mbox{diag}\biggl({4\over 5}, {2\over 5}, 
0, -{2\over 5}, -{4\over 5}\biggr)
+ \mbox{permutations}\quad\mbox{for}\quad SU(5).
\end{eqnarray}
Since $\theta_i$ is a module of $2\pi$, the configuration for $SU(5)$ 
is equivalent to 
\begin{equation}
\vev{\Phi} ={\pi\over{gL}}\mbox{diag}\biggl({4\over 5}, {2\over 5}, 0, 
{8\over 5}, {6\over 5}\biggr) + \mbox{permutations}.
\end{equation}
\par
The configuration (\ref{config}) breaks the $SU(N)$ gauge symmetry
maximally to $U(1)^{N-1}$. It should be noted that this does not depend on  
the number of flavor $N_F$ and the supersymmetry breaking 
parameter $\beta$, which is the boundary condition of the 
fields $\lambda, \phi$ for the $S^1$ direction. We 
have arrived at the conclusion that in our model
the $SU(N)$ gauge symmetry is broken dynamically to its maximal commutative 
subgroup, {\it i.e.}, $U(1)^{N-1}$, independent of $\beta$ and
$N_F$. This is very 
different from the nonsupersymmetric gauge theories, in which
the symmetry breaking patterns depend on the 
boundary conditions of the matter fields \cite{hohosotani, ho}.
\par
We also depict potential energies for the possible gauge symmetry
breaking patterns of $SU(3), SU(5)$ in Figs. $1$ and $2$. We compare 
the potential energy for each pattern \cite{hohosotani} given by
\begin{eqnarray}
SU(3)&\rightarrow&\left\{\begin{array}{crl}
\mbox{A}:&\hspace{1.9cm}U(1)\times U(1)~\cdots &{\pi\over 3}(2, 0, -2)+
\mbox{permutations},\\
\mbox{B}:&SU(2)\times U(1)~\cdots &{\pi\over 3}(1, 1, -2)+
\mbox{permutations},\\
\mbox{C}:&SU(3)~\cdots & {\pi\over 3}(0, 0, 0),\\
\end{array}\right.\label{patternsa}\\
SU(5)&\rightarrow&\left\{\begin{array}{crl}
\mbox{A}:&U(1)^4~\cdots &{\pi\over 5}(4, 2, 0, 8, 6)+\mbox{permutations},\\
\mbox{B}:&SU(2)^2\times U(1)^2~\cdots &
{\pi\over 5}(0, 3, 3, 7, 7)+\mbox{permutations},\\
\mbox{C}:&SU(3)\times SU(2)\times U(1)~\cdots & 
{\pi\over 5}(0, 0, 0, 1, 1)+\mbox{permutations},\\
\mbox{D}:&SU(4)\times U(1)~\cdots & 
{\pi\over 5}(1, 1, 1, 1, 6)+\mbox{permutations},\\
\mbox{E}:&SU(5)~\cdots & 
{\pi\over 5}(0, 0, 0, 0, 0).\label{patternsb}\\
\end{array}\right.
\end{eqnarray}
It is clear from the figures that that the lowest energy of the potential 
is always given by the case of the maximal breaking of the original 
gauge symmetry and that the boundary condition $\beta$
never affects the symmetry breaking patterns.
\par
The potential energy at the minimum\footnote{The
vacuum energy may be estimated by taking account of
the vacuum expectation
values of the squark $\vev{\phi}\equiv a \in{\bf C}$ in addition
to $\vev{\Phi}$. Originally, this corresponds to
the flat direction satisfying the $D$-term condition. The
flat direction may be lifted in our case due to $\beta$. The model we are
considering has essentially two kinds of order 
parameters $\vev{\Phi}$ and $\vev{\phi}$, though the leading 
contribution to the potential may be 
dominated by $\vev{\Phi}$. 
This is inferred from the fact that the effective 
potential (\ref{totpot}) does not depend on the gauge coupling 
at the one-loop level.
We are focusing here on
the dynamics of the nonintegrable phases.} is calculated as
\begin{equation}
V(\theta)=(2N_F-2){{4\pi^2{\bar\beta}^2}\over {L^4}}
\biggl[{N\over{12}}({\bar\beta}-1)^2+{{N(N-1)}\over{12}}({\bar\beta}^2+1)
-{1\over 3}(N-1)(N-2)
\biggr],
\end{equation}
where we have rescaled $\beta$ as $\beta=2\pi\bar\beta$.
It is obvious to see that the energy is positive for $SU(2)$. If the 
gauge group becomes larger than $SU(2)$, the sign of the
energy depends on the values of $N$ and $\beta$. Let us define 
\begin{equation}
D(N)\equiv N({\bar\beta}-1)^2+N(N-1)({\bar\beta}^2+1)-4(N-1)(N-2).
\end{equation}
Then, we find that 
\begin{eqnarray}
V(\theta)&\geq& 0\quad\mbox{for}\quad 2\leq N\leq N_+\equiv 
{{6-{\bar\beta}+\sqrt{9{\bar\beta}^2-12{\bar\beta}+12}}\over
{3-{\bar\beta}^2}}<4,\nonumber\\
V(\theta)&<& 0\quad \mbox{for}\quad 4\leq N.
\end{eqnarray}
The zero energy can be realized only by $N=3$ 
with $\bar\beta={1\over 3}$. The potential energy is positive definite for the
other values of $\beta$ when $N=3$.
\par
\section{Mass Spectrum and Low-energy Effective Theory}
\subsection{Higgs Scalar}
Let us study the mass of the Higgs 
scalar $\Phi\equiv A_3$, which is
originally the component gauge field for the $S^1$ direction.  
The mass term for the Higgs scalar, which is zero at the tree level, is 
generated 
through the quantum corrections in the extra dimensions \cite{hosotania}.
After the compactification is carried out by integrating 
the coordinate of $S^1$, the lowest Kaluza-Klein mode of $\Phi$ behaves as 
the adjoint Higgs scalar, which transforms as the adjoint representation
under the gauge group. 
\par
Let us now evaluate the mass term for the  
Higgs scalar. It is 
given by estimating the second derivative of the
effective potential (\ref{potformula}) at the 
absolute minimum (\ref{sol}). Alternatively, after taking the second
derivative of Eq. (\ref{potanother}) with respect to $\theta_i$, one can use
$\sum_{n=1}^{\infty}{1\over {n^2}}\cos(nt)={1\over 4}(t-\pi)^2
-{{\pi^2}\over{12}}$, which is also obtained by taking the second derivative
of Eq. (\ref{formula}) with respect to $t$. In this approach, it is 
helpful to notice that
\begin{equation}
\sum_{n=1}^{\infty}{1\over n^2}
\biggl(2\cos[nx]-\cos[n(x-\beta)]-\cos[n(x+\beta)]\biggr)
=-{\beta^2\over 2},
\end{equation}
which is independent of $x$. The two approaches give the same result, 
as they should. Then, we obtain
\begin{equation}
\mbox{mass term}\equiv\half~\theta_i~(M^{\Phi})_{ij}~\theta_j\quad
(i, j=1,\cdots, N-1),
\end{equation}
where 
\begin{equation}
(M^{\Phi})_{ij}\equiv
{{\del^2V(\theta)}\over {\del\theta_i\del\theta_j}}=
{{(2N_F-2)}\over{\pi^2L^4}}{\beta^2\over 4}\times 2N\times
\left(\begin{array}{ccccc}
2&1&\cdots&\cdots& 1\\
1&2& & &\vdots\\
\vdots& &\ddots & &\vdots\\
\vdots& & & \ddots&\vdots\\
1&\cdots&\cdots&\cdots&2\\
\end{array}\right).
\label{matrix}
\end{equation}
The indices $i, j$ run from $1$ to $N-1$. The matrix in Eq. (\ref{matrix})
is the same as the one in Eq. (\ref{stationary}).
\par
In order to study the mass of the Higgs 
scalar more clearly, let us change
the variable $\theta_i$ to another variable.
To this end, let us define
\begin{equation}
\vev{\Phi}\equiv{1\over{gL}}\sum_{m=1}^{N-1}v_mH_m,
\label{cartandecom}
\end{equation}
where $H_m (m=1, 2, \cdots, N-1)$ is the diagonal generator 
of the Cartan subalgebra of $SU(N)$ and is a $N\times N$ matrix. 
$v_m$ is a real parameter. We choose the form 
\begin{equation}
(H_m)_{ij}={1\over{\sqrt{2m(m+1)}}}\Bigl(
\sum_{k=1}^m\delta_{i, k}\delta_{j, k}
-m\delta_{i,m+1}\delta_{j, m+1}\Bigr).
\end{equation}
Then, the $\theta_i$'s in Eq. (\ref{background})
are related with $v_m$ by 
\begin{eqnarray}
\theta_1&=&~~{v_1\over 2}+{v_2\over{2\sqrt{3}}}
+\cdots+{{v_m}\over{\sqrt{2m(m+1)}}}+\cdots +{{v_{N-1}\over
\sqrt{2N(N-1)}}},    \nonumber\\
\theta_2&=&-{v_1\over 2}+{v_2\over{2\sqrt{3}}}
+\cdots\cdots\hspace{3.32cm}+{{v_{N-1}\over\sqrt{2N(N-1)}}},\nonumber\\
\theta_3&=&\hspace{1.2cm}-{v_2\over{\sqrt{3}}}
+{{v_3}\over{2\sqrt{6}}}+\cdots\cdots 
\hspace{2.0cm}+{{v_{N-1}\over\sqrt{2N(N-1)}}}, \nonumber\\
\vdots& &\hspace{4.5cm}\vdots     \nonumber\\
\theta_{m+1}&=&\hspace{3.5cm}{-m{v_m}\over{\sqrt{2m(m+1)}}}+
\cdots +{{v_{N-1}\over\sqrt{2N(N-1)}}},\nonumber\\
\vdots& &\hspace{4.5cm}\vdots      \nonumber\\
\theta_{N-1}&=&\hspace{3.5cm}
{{-(N-2)v_{N-2}}\over{\sqrt{2(N-1)(N-2)}}}
+{{v_{N-1}}\over{\sqrt{2N(N-1)}}}.
\label{relation}
\end{eqnarray}
$\theta_{N}=-\sum_{i=1}^{N-1}\theta_i$ is 
given by $-(N-1)v_{N-1}/\sqrt{2N(N-1)}$.
It follows from Eq. (\ref{cartandecom}) that $\vev{\Phi^m}=v_m/gL$.
We obtain an equation which relates $\theta_i$ with $\vev{\Phi^m}$ 
as follows:
\begin{equation}
\left(\begin{array}{c}
\theta_1\\
\theta_2\\
\vdots\\
\theta_{N-1}
\end{array}\right)=gL\times B\left(\begin{array}{c}
\vev{\Phi^1}\\
\vev{\Phi^2}\\
\vdots\\
\vev{\Phi^{N-1}}
\end{array}\right),
\end{equation}
where $B$ is a $(N-1)\times (N-1)$ matrix given by
\begin{equation}
B=\left(\begin{array}{ccccccc}
\half & {1\over{2\sqrt{3}}}&\cdots & \cdots & {1\over {\sqrt{2m(m+1)}}}& 
\cdots &{1\over{\sqrt{2N(N-1)}}} \\
-\half & {1\over{2\sqrt{3}}}& \cdots &\cdots & \cdots& \cdots &
{1\over{\sqrt{2N(N-1)}}} \\
0 & -{1\over{\sqrt{3}}}& {1\over{2\sqrt{6}}}&\cdots & \cdots& \cdots &
{1\over{\sqrt{2N(N-1)}}} \\
\vdots & \vdots&\vdots & \vdots&\vdots& \vdots &\vdots \\
0& \cdots& \cdots& 0& -{m\over{\sqrt{2m(m+1)}}}& \cdots 
& {1\over{\sqrt{2N(N-1)}}}\\
\vdots & \vdots& \vdots & \vdots& \vdots &\vdots \\
0 & \cdots& \cdots& \cdots & 0 &{{-(N-2)}\over{\sqrt{2N(N-1)}}} 
& {1\over{\sqrt{2N(N-1)}}}\\
\end{array}\right).
\end{equation}
The row vector in the matrix $B$ is just
the weight vector $\nu^i (i=1\sim N-1)$ in the fundamental
representation of $SU(N)$. Thus, we obtain
\footnote{One can diagonalize the matrix $(M^{\Phi})_{ij}$ by a 
real symmetric matrix $U$. The eigenvalues 
are $1 [(N-2)$-degeneracy] and $N$. If we rescale $\theta_i$ as 
$\theta_i\rightarrow \theta_i/\sqrt{2} (i=1, \cdots, N-2)$ 
and $\theta_{N-1}\rightarrow \theta_{N-1}/\sqrt{2N}$, then the matrix
$U$ becomes $B$ in the text. Accordingly, all the eigenvalues 
are scaled to be $N/2$. This also 
implies that the inverse of $B$ is given by taking the transposition of 
$B$ and multiplying the first $N-2$ numbers of the row vectors by $2$
and the last $(N-1)$th row vector by $2N$.}
\begin{equation}
\mbox{mass term}=
\half~(gL)^2\theta_i~(M^{\Phi})_{ij}~\theta_j
=\half~\Phi^i~[B^T
M^{\Phi}~B]_{ij}\Phi^j,
\end{equation}
where 
\begin{equation}
B^T~M^{\Phi}~B
={{(N_F-1)}\over{\pi^2}}{{g^2\beta^2}\over L^2}
{N\over 2}{\bf 1}_{(N-1)\times (N-1)}.
\end{equation}
We have found that the masses for the Higgs 
scalars $\Phi^1, \Phi^2, \cdots, \Phi^{N-1}$ are all degenerate and are 
given by
\begin{equation}
m_{\Phi}^2
={{(N_F-1)}\over{\pi^2}}{{g^2\beta^2}\over L^2}
{N\over 2}.
\label{adjhiggs}
\end{equation}
\par
The mass term respects the residual gauge symmetry $U(1)^{N-1}$ and  
explicitly depends on the gauge coupling $g$, the supersymmetry 
breaking parameter $\beta$, and the number of flavor $N_F$.
The Higgs boson mass $m_{\Phi}^2$ suffers 
from a correction of $O(1/L^2)$. If we would consider 
massive adjoint matter instead of a massless one, the mass
of the Higgs scalar would have a ``Boltzman factor'' such as $\e^{-mL}$
for large $mL$, where 
$m$ is the mass of the adjoint matter \cite{lim}. 
\par
If we take $\beta=0$, the supersymmetry is restored, and all the 
components of the adjoint Higgs scalar become massless. 
This reflects the fact that the effective potential is flat
for all directions with respect to $\Phi^a (a=1, \cdots, N^2-1)$.
Since the
origin of the 
mass of the Higgs scalar is the quantum corrections in the
extra dimensions, even though the 
gauge symmetry is not broken dynamically, nonvanishing mass terms 
respecting the unbroken gauge symmetry may appear.
\par
\subsection{Mass Spectrum in Three Dimensions}
Since the $SU(N)$ gauge symmetry is broken to $U(1)^{N-1}$ 
in our model, the gauge
boson $(A_{\mu})$, gaugino $(\lambda)$, squark $(\phi)$, and 
quark $(q)$ become massive. 
In order to obtain the mass terms at the tree level, it may be 
convenient to use the parametrization (\ref{cartandecom}) and to 
expand the fields as
\begin{equation}
F=F^mH_m+F^{(\alpha)}E_{\alpha},\quad F\equiv A_{\mu},~\lambda,~\phi,~q.
\label{decomp}
\end{equation}
$E_{\alpha}$ is the raising and lowering operator corresponding 
to the roots $\alpha$ of the Lie algebra.
Using the commutation relations
\begin{equation}
[H_m,~H_n]=0,\quad [H_m,~E_{\pm\alpha}]=\pm\alpha_mE_{\pm\alpha}~ 
(E_{-\alpha}=E_{\alpha}^{\dagger}),\quad
\mbox{tr}(E_{\alpha}E_{\beta}^{\dagger})=\half\delta_{\alpha\beta},
\end{equation}
we find that
\begin{eqnarray}
-g^2~\mbox{tr}([\vev{\Phi},~A_{\mu}])^2&=&
{1\over {2L^2}}\sum_{\alpha}(v_m\alpha_m)^2\abs{A^{(\alpha)}}^2,\nonumber\\
2g^2~\mbox{tr}([\vev{\Phi},~\phi^{\dagger}][\vev{\Phi},~\phi])&=&
-{1\over{L^2}}\sum_{\alpha}(v_m\alpha_m)^2\abs{\phi^{(\alpha)}}^2,\nonumber\\
g~\mbox{tr}(-\lambda\sigma^3[\vev{\Phi},~{\bar\lambda}]+[\vev{\Phi},~\lambda]
\sigma^3{\bar\lambda})&=&{1\over L}\sum_{\alpha}(v_m\alpha_m)
\lambda^{(\alpha)}\sigma^3{\bar\lambda}^{(\alpha)},\nonumber\\
g~\mbox{tr}(-\lambda\sigma^3[\vev{\Phi},~{\bar q}]+[\vev{\Phi},~q]
\sigma^3{\bar q})&=&{1\over L}\sum_{\alpha}(v_m\alpha_m)
q^{(\alpha)}\sigma^3{\bar q}^{(\alpha)},
\end{eqnarray}
where $v_m$ is obtained by Eq. (\ref{config}) through Eq. (\ref{relation}).
\par
The field $F^{(\alpha)}$ behaves as ``charged'' field under the residual
gauge symmetry $U(1)^{N-1}$. If we decompose the Weyl spinor 
in four dimensions
into the one in three dimensions, we have
\begin{equation}
\psi^{(\alpha)}\sigma^3{\bar\psi}^{(\alpha)}\rightarrow\half 
({\bar\psi}_1^{(\alpha)}\psi_1^{(\alpha)}
+{\bar\psi}_2^{(\alpha)}\psi_2^{(\alpha)}), \quad 
(\psi^{(\alpha)}=\lambda^{(\alpha)}, q^{(\alpha)}).
\end{equation}
\par
Let us proceed to the mass spectrum in 
three dimensions by integrating
the coordinate of $S^1$.
The Kaluza-Klein mass is generated through the kinetic
term for the compactified direction. In $S^1$ compactification, there
appears no Kaluza-Klein mass for the Higgs scalar $\Phi$ because 
there is no coupling of $\mbox{tr}(F_{ab})^2$, where $a, b$ stand for
the compactified coordinates. By straightforward calculations, we obtain
the mass terms for gauge boson, gaugino, squark, quark, and
the Higgs scalar. These are
summarized as follows:
\begin{eqnarray}
A_{\hmu}&=&\left\{\begin{array}{ll}
A^m_{(n)\mu} ~~\cdots         & ({{2\pi n}\over L})^2, \\
A^{(\alpha)}_{(n)\mu}~~\cdots & \Bigl({{2\pi n}\over L}
+{{(v_m\alpha_m)}\over L}\Bigr)^2, \label{gauge}\\
\end{array}\right. \\  
\Phi(\equiv A_3)&=&\left\{\begin{array}{ll} 
\Phi^m_{(n)} ~~\cdots       & {{(N_F-1)g^2\beta^2 N}\over {2\pi^2 L^2}},\\
\Phi^{(\alpha)}_{(n)}~~\cdots & \mbox{massless},\label{adjoint} \\
\end{array}\right.\\
\lambda &=& \left\{\begin{array}{ll}
\lambda_{(n)i}^m ~~\cdots         & {{2\pi}\over L}(n+{\beta\over {2\pi}}), \\
\lambda_{(n)i}^{(\alpha)}~~\cdots & {{2\pi}\over L}
(n+{\beta\over{2\pi}})-{{(v_m\alpha_m)}\over L}, \label{gaugino}\\
\end{array}\right. \\
\phi &=& \left\{\begin{array}{ll}
\phi^m_{(n)}~~\cdots & 
\Bigl({{2\pi n}\over L}(n+{\beta\over{2\pi}})\Bigr)^2, \\
\phi^{(\alpha)}_{(n)}~~\cdots & \Bigl({{2\pi n}\over L}
(n+{\beta\over{2\pi}})+{{(v_m\alpha_m)}\over L}\Bigr)^2, \label{squark}\\
\end{array}\right. \\
q &=& \left\{\begin{array}{ll}
q_{(n)i}^m~~\cdots          & {{2\pi n}\over L}, \\
q_{(n)i}^{(\alpha)}~~\cdots & {{2\pi n}\over L}-{{(v_m\alpha_m)}\over L}. 
\label{quark}\\
\end{array}\right.
\end{eqnarray}
The Kaluza-Klein mode is denoted by $(n)$. The index $i(=1, 2)$
stands for the three-dimensional Majorana spinor. 
\par
We observe that all the mass terms
are proportional to the compactification scale $1/L$ 
and do not depend on the gauge coupling 
constant $g$ except for the mass of the Higgs scalar $\Phi$. 
The independence of the gauge coupling in the form of
$(v_m\alpha_m)/L$ shows that  
the mass generated  
through the dynamical gauge symmetry breaking 
is the leading effect besides the Kaluza-Klein mode even though 
the dynamics itself is caused by the quantum corrections 
in the extra dimensions.
The dependence of the Higgs mass on the gauge coupling means that
the mass is generated by the quantum effects.
Both components $F^m$ and 
$F^{(\alpha)}$ acquire the Kaluza-Klein mass. 
The zero modes for $\lambda_{(n)i}^m$ and $\phi_{(n)}^m$ are 
removed by the Scherk-Schwarz mechanism.
Let us note that 
$\lambda_{(n=0)}^{(\alpha)}$ and $\phi_{(n=0)}^{(\alpha)}$
can be massless for special values 
of $\beta=\pm (v_m\alpha_m)/2\pi$, for example, when  
the gauge symmetry breaking occurs. This is one of special features
resulting from the existence of the supersymmetry breaking 
parameter $\beta$.
\par
\subsection{Low-energy Effective Theory}
In this section we discuss low-energy effective theories 
in three dimensions. We are, in particular, interested in the low-energy
effective theory, in which relevant energy scale is much smaller than the
compactification scale $1/L$. If we take $L$ to be small, it 
corresponds to small extra dimensions. 
\par
The particle whose mass scale is proportional to $1/L$ becomes superheavy 
if we take the small extra dimensions, and it is 
decoupled from low-energy physics.
Only massless particles survive at low energies, so that
the low-energy effective theory consists
of them. This means that all the effects of the quantum corrections in
the extra dimensions disappear
in low energies as long as we take the limit of the small extra dimensions.
Nevertheless, the nontrivial limit of the small extra dimensions is possible
thanks to the supersymmetry breaking parameter $\beta$ of the Scherk-Schwarz 
mechanism. 
\par
Let us first consider the naive limit of small extra dimensions.
As we have mentioned above, only the massless
mode survives at low energies. The massless particles 
are $A_{(n=0)\mu}^m, q^m_{(n=0)i}$ and $\Phi_{(n=0)}^{(\alpha)}$, which 
consist of 
the low-energy effective theory. This is summarized 
in (i) of Table I. If we do not take the quantum corrections
in the extra dimensions into account and consider the small extra 
dimensions, the particle contents in the low-energy effective theory
are given by the trivial compactification. 
This is also listed in (ii) of Table I. Let us note that the 
superparticles, gaugino ($\lambda^m$) and squark ($\phi^m$), do not 
have massless modes due to the supersymmetry
breaking parameter $\beta$, so that they do not appear in the
low-energy effective theories.
\par
Let us consider the limit by which some particle masses remain finite
even in the limit of small extra dimensions. Suppose that $\bar\beta$
and $L$ are the same order, and we take $\bar\beta, L\rightarrow 0$, keeping
the ratio of $\bar\beta$ and $L$ finite:
\begin{equation}
{{\bar\beta}\over L}=\mbox{fixed~~~~as} \quad {\bar\beta},~L \rightarrow 0,
\label{limitb}
\end{equation}
where $\beta\equiv 2\pi\bar\beta$. 
In this limit, as seen 
from Eq. (\ref{adjhiggs}), the mass of the the 
Higgs scalar 
$m_{\Phi}^2$ is finite, and the Higgs scalar survives at low energies.
The limit also makes the masses for $\lambda_{(n=0)i}^m$ and 
$\phi_{(n=0)}^m$ finite, so that they, superparticles, also 
take part in the low-energy
physics. 
The limit gives us a technique for generating mass in low-energy
effective theory through compactification.
Let us note that 
as long as $(v_m\alpha_m)/L$ is nonzero, the 
masses for $\lambda_{(n=0)i}^{(\alpha)}$ and
$\phi_{(n=0)}^{(\alpha)}$ become superheavy even in this 
nontrivial limit. The particle 
contents in the low-energy effective theory are summarized in (iii) of 
Table I. 
\begin{center}
Table I
\end{center}
\begin{center}
\begin{tabular}{|c|l|l|}\hline
 & Supersymmetric Yang-Mills theory& \\ 
 & with massless adjoint matter    &  \\ \hline
Limit& Particle contents& Symmetry    \\ \hline
(i) & $A_{(n=0)\mu}^{m}, \hspace{2.45cm}\Phi^{(\alpha)}_{(n=0)}, 
q_{(n=0)i}^m$ &
$U(1)^{N-1}$\\\hline
(ii) & $A_{(n=0)\mu}^{m}A_{(n=0)}^{(\alpha)}, 
\Phi_{(n=0)}^{m}, \Phi_{(n=0)}^{(\alpha)}, q_{(n=0)i}^m, 
q_{(n=0)i}^{(\alpha)}$ &$SU(N)$\\\hline
(iii)
& $A_{(n=0)\mu}^{m},\hspace{1.15cm}\Phi^{m}_{(n=0)}, 
\Phi^{(\alpha)}_{(n=0)}, q_{(n=0)i}^m, 
\hspace{1.10cm}\lambda_{(n=0)i}^{m}, \phi_{(n=0)}^m$
&$U(1)^{N-1}$\\\hline
\end{tabular}
\end{center}
\begin{flushleft}
Table I: Particle contents in the low-energy effective theories
of $SU(N)$ supersymmetric Yang-Mills theory with massless adjoint matter.
(i) stands for the naive 
limit of $L\rightarrow 0$ for the theory taking the quantum corrections 
in the extra dimensions into account.
(ii) is the trivial compactification and (iii) represents the
nontrivial limit defined by Eq. (\ref{limitb}).
\end{flushleft}
\par
There is no supersymmetry in the effective
theories because of the nontrivial phase $\beta$, and the gauge symmetry
is broken to $U(1)^{N-1}$ through the dynamics of the nonintegrable
phases except in the case of (ii) in Table I.
We observe that 
the superparticles $\lambda^m_{(n=0)}, \phi^m_{(n=0)}$ come into 
play in the low-energy physics
in the case of the limit (\ref{limitb}). The quantum corrections in 
the extra dimensions become manifest in the low-energy physics by
the mass of the adjoint Higgs scalar $\Phi^m_{(n=0)}$. It should be
noted that this is possible due to the existence of the unique parameter
$\beta$ associated with the 
Scherk-Schwarz mechanism of supersymmetry breaking.
\par
The above discussion on low-energy effective theory 
is based on classical considerations of the mass 
terms (except for the Higgs scalar $\Phi$) in the previous subsection. 
In general if we take into account quantum 
corrections for the mass terms, they 
suffer from the correction of order $O(1/L)$, like the Higgs scalar.
This may modify the massless 
modes at the tree level and make them superheavy. They are 
decoupled from low-energy physics.
Therefore, the effective theory may be 
different from the one obtained from the classical consideration.
If $L$ is large, the quantum corrections are suppressed, so that the 
classical consideration may be a good approximation for the effective theory.
The particle can still remain massless even after taking into account the
quantum corrections if they are the Nambu-Goldstone 
boson, for example, associated with the breakdown of symmetry.
And the limit defined in Eq. (\ref{limitb}) is an example of obtaining 
massive particles at low energies.
\par
\section{Conclusions} 
We have studied the dynamics of the nonintegrable phases
in $SU(N)$ supersymmetric Yang-Mills theory with $N_F$ numbers of
massless adjoint matter. We have resorted to the 
Scherk-Schwarz mechanism, by which supersymmetry is broken softly, in 
order to obtain the nonvanishing 
effective potential for the phases in perturbation theory. 
\par
We have found that the $SU(N)$ gauge symmetry 
is broken dynamically to its maximal
commutative subgroup, {\it i.e.}, $U(1)^{N-1}$. 
This result does not depend on  
the values of the supersymmetry breaking parameter $\beta$, which is 
the boundary condition of the fields $\lambda, \phi$ 
for the $S^1$ direction. This is remarkable if we compare our model with
nonsupersymmetric gauge theories, in which rich 
symmetry breaking patterns are possible, depending
on the values of the boundary conditions of the fields.
\par
We have obtained the mass of the  
Higgs scalar $\Phi$, which is originally the
component gauge field for the compactified direction. The mass
is generated through the quantum corrections in the extra dimensions. 
The mass term respects the residual gauge 
symmetry $U(1)^{N-1}$ and explicitly depends on the gauge 
coupling $g$, the supersymmetry breaking parameter $\beta$, and 
the number of flavor $N_F$. The Higgs boson mass suffers from a correction 
of $O(1/L^2)$.
\par
We have also obtained the mass spectrum in three dimensions and
discussed the low-energy effective theory in the limit of the small
extra dimensions. All the effects of the extra dimensions are decoupled 
from low energies in the naive limit of $L\rightarrow 0$ since the 
relevant mass scale is given by the compactification scale $1/L$
as shown in Eqs. (\ref{gauge})-(\ref{quark}).
We have considered the nontrivial
limit defined by Eq. (\ref{limitb}). The nontrivial limit is possible
thanks to the supersymmetry breaking parameter $\beta$ of the
Scherk-Schwarz mechanism. The mass of the Higgs scalar
becomes finite in the limit, and the Higgs scalar 
survives at low energies. This implies that the quantum corrections
in the extra dimensions 
become manifest in the low-energy physics even in the
limit of the small extra dimensions.
The limit also makes the masses of the gaugino $\lambda$
and squark $\phi$ finite, so that the superparticles 
come into play at low energies.  
\par
Concerning the limit defined by Eq. (\ref{limitb}), let us 
comment on the gauge coupling constant in the low-energy 
effective theory. If we 
start with a space-time $M^{D-m(=4)}\otimes T^m (m\mbox{-torus})$ and 
assume that the size of the extra dimensions is equal to $L^m$, then the 
dimensionless gauge coupling constant
in $M^{D-m}$ is given by ${\tilde g}=g/L^{m/2}$. The 
trilinear and quartic coupling constants arising from the covariant 
derivative have the 
form $g(2\pi n/L +\beta/L)/L^{m/2}={\tilde g}(2\pi n/L + \beta/L)$ and 
$g^2/L^m={\tilde g}^2$, respectively. And the mass of the Higgs
scalar is scaled on the dimensional ground 
as $m_{\Phi}^2\sim g^2L^2\beta^2/L^D={\tilde g}^2(\beta/L)^2$, where $D=4+m$. 
As long as $\tilde g$ is 
finite and we take the limit (\ref{limitb}), the coupling constants and the 
mass of the Higgs scalar are finite at low energies.
\par
It may be interesting to introduce
the matter fields belonging to the fundamental representation under the 
$SU(N)$ gauge
group in addition to the adjoint matter. Then, unlike the 
present case, we may expect rich patterns of
gauge symmetry breaking and realistic low-energy
effective theories. These are under investigation.
\par
\vskip 2cm
\begin{center}
{\bf Acknowledgments}
\end{center}
The author would like to thank P. Ferreira, M. Sakamoto, and T. Tsukioka
for valuable discussions and the Dublin Institute for Advanced Study
for warm hospitality.

\newpage
\vspace*{3cm}
\begin{center}
\epsfxsize=16cm\epsffile{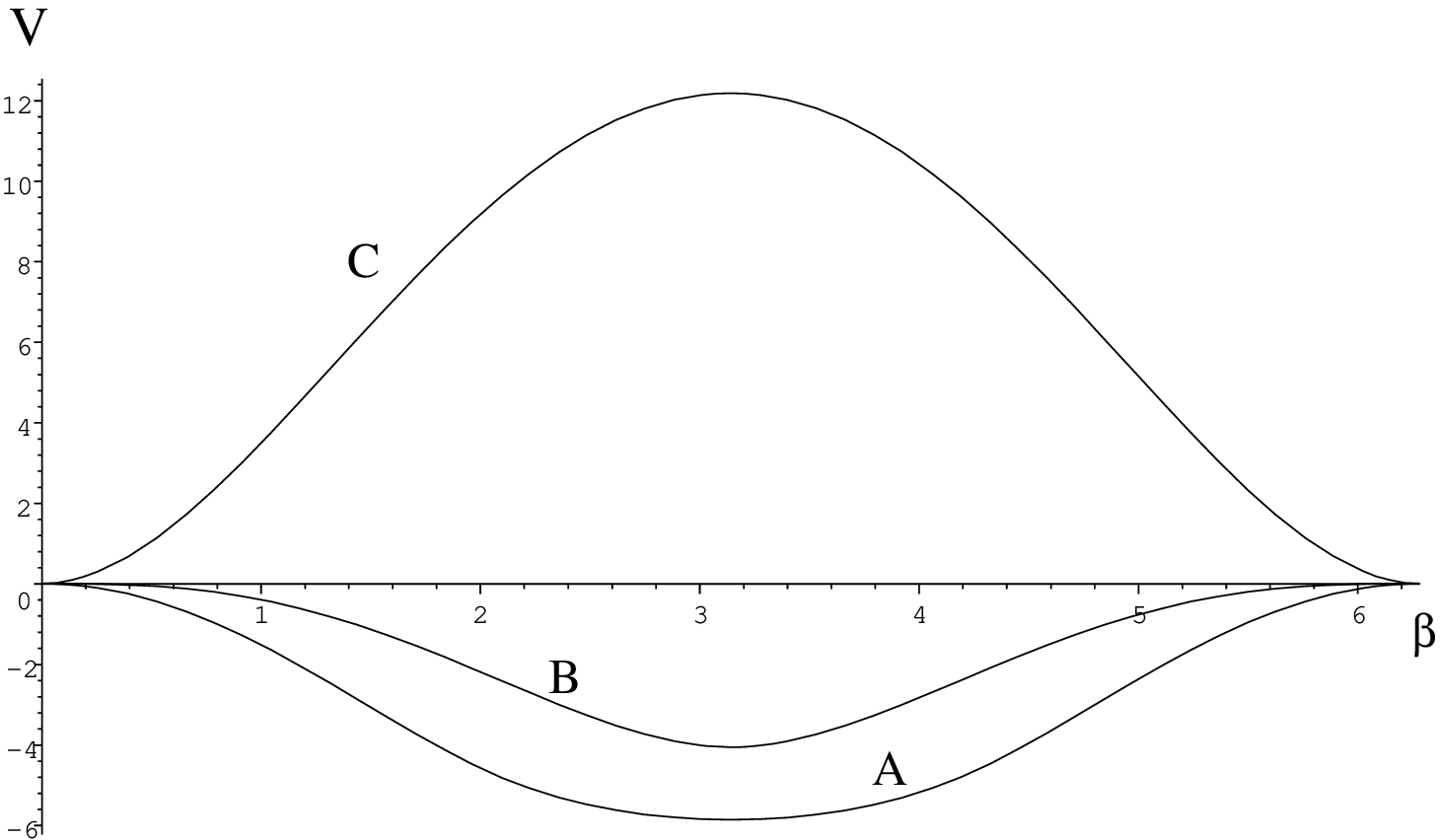}
\end{center}
\smallskip
FIG. $1$. Potential energies for the gauge symmetry 
breaking patters of $SU(3)$, Eq. (\ref{patternsa}). The horizontal 
axes stand for the supersymmetry breaking parameter $\beta$
of the Scherk-Schwarz mechanism. In calculating the potential energy
numerically, we have ignored 
the factor as well as the terms which do not depend
on $\theta_i$ in the 
effective potential (\ref{potanother}).
\newpage
\vspace*{3cm}
\begin{center}
\epsfxsize=16cm\epsffile{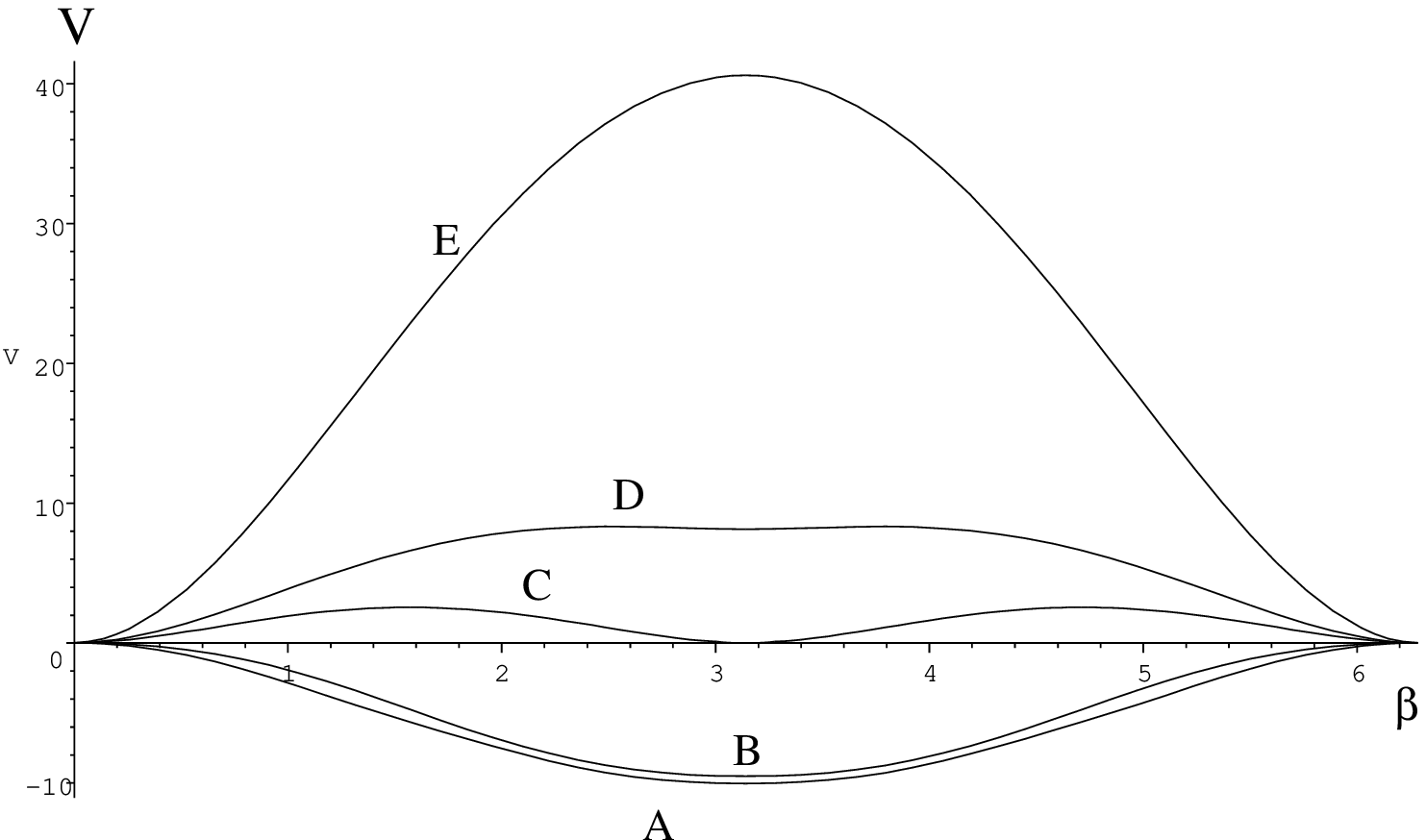}
\end{center}
\smallskip
FIG. $2$. Potential energies for the gauge symmetry 
breaking patters of $SU(5)$, Eq. (\ref{patternsb}). The 
horizontal axes stand for the supersymmetry breaking parameter $\beta$
of the Scherk-Schwarz mechanism. In calculating the potential energy
numerically, we have ignored 
the factor as well as the terms which do not depend
on $\theta_i$ in the effective potential (\ref{potanother}).
\end{document}